\documentclass[pr,aps,praps,amsbsy]{revtex4}
\usepackage{graphics}

\begin{document}

\title{\normalsize{\bf{Device on  basis of a bent crystal with variable curvature 
for particle beams steering in accelerators.
 }}}

\author{  A.G.Afonin, V.T.Baranov, M.K.Bulgakov, Yu.A.Chesnokov,  P.N.Chirkov,  E.V.Lobanova, I.S.Lobanov, 
A.N.Lunkov,  V.A.Maisheev, I.V.Poluektov,  Yu.E.Sandomirskiy, I.A.Yazynin.
\\
  \it{ Institute for High Energy Physics, 142281, Protvino, Russia }}

\begin{abstract}
Recently it was proposed  to apply a bent single crystal with decreasing curvature instead of uniform bending  for improvement of extraction and 
collimation of a circulating beam in  particle accelerators. In the given paper created crystal devices with a variable curvature, realizing this 
idea are described. Results of measurement of curvature along a crystal plate are informed. It is shown, that with the help of the developed 
devices it is possible to carry out also high energy beam focusing. The mathematical description of this process is proposed.
\end{abstract}

\maketitle


\section{Introduction}
    At present, the collimation and extraction  of a circulating beam using coherent phenomena in oriented single crystals are examined 
at several large accelerators. Pioneering works \cite{1,2,3,4} at the U-70 IHEP accelerator indicate that, in short bent silicon crystals, 
channeling can increase the efficiency of the extraction and collimation of the beam up to 85\%. 
This possibility has been confirmed at the SPS collider (CERN) \cite{5} and at the Tevatron (Fermilab) \cite{6}. In view of 
the commissioning of the Large Hadron Collider (LHC) and the problem of an increase in its luminosity, 
the problem of improving the beam collimation efficiency is of particular importance \cite{7}.    In \cite{8} it was proposed  
to apply the bent single crystal with decreasing  curvature instead of uniform bending  for improvement of extraction and 
collimation of a circulating beam in  particle accelerators. Presence of decreasing  curvature in a crystal leads to  
suppression of dechanneling  and thus way improves parameters of extraction/collimation of a circulated beam.  
In the given work the created devices with variable curvature are described. Results of measurement of curvature 
along a crystal plate are informed. It is shown, that with the help of the developed devices it is possible to carry out 
also high energy beam focusing. 

\section{The scheme  of  crystal bending  with variable curvature}

 For realization of  effective extraction  of a proton beam from U-70 \cite{1,2,3,4} it was necessary to find  a challenge decision how 
to  bend the short silicon crystals on small angles. With this purpose a way \cite{1,4}  of bending  of a crystal as narrow strips  by length  
of  $\sim$2 mm along the beam and 40 mm in height has been developed  based on use of anisotropic properties of crystal lattices. 
From the theory of elasticity it is known, that at a bend of a crystal plate in  longitudinal direction in an orthogonal  direction there are 
the deformations accepting saddle or barrel  form depending on concrete anisotropic properties of a material and orientation 
of a crystal (see for example \cite{9}, page 85). In silicon single crystals the greatest orthogonal  deformations are formed at orientation (111) and 
get the saddle  form (see Fig. 1à). Thus the bend of a crystal on height on  $\sim$100 mrad angle   provides in an orthogonal  direction 
the bending angle of about 1 mrad sufficient for particle extraction from accelerator \cite{4}.

     The subsequent precision experiments executed on beam line  Í8 in northern CERN area with microstrip detectors, 
have confirmed, that similar crystals effectively deflect  400 GeV protons and are uniformly bent  \cite{10}. We have applied 
the same method of a bend of a silicon strip to a nonuniform bend of a crystal, but used a plate of trapezoidal  cross- section (Fig. 1b). 
In this case due to falling down thickness of a strip different mechanical tensions  in crystal  material along the   horizontal direction 
(on coordinate z) are created, that leads to decreasing the bending  curvature as approaching the narrow side of a trapeze. 
Empirically picked up tangent   of an inclination of a lateral face of a trapeze has allowed to reach  the necessary gradient of curvature. 

\section{Measurement of the bending shape  of a crystal}

     In Fig. 2a   is shown a photo of proposed crystal  device. The silicon crystal plate  1 of trapezoidal cross- section 
is bent in a longitudinal direction by the metal holder 2. With the help of the screw 3 it is provided vertical, on coordinate y 
(longitudinal) bend of a crystal plate on 20 mrad. Screws 4 eliminate possible  twisting a plate on height (eliminate "effect of a propeller" ).
Such design of the bending holder for the first time has 
been offered by Yu.Chesnokov in [1] and then modified in 
several publications by various researchers.

\begin{figure}
\scalebox{0.6}{\includegraphics{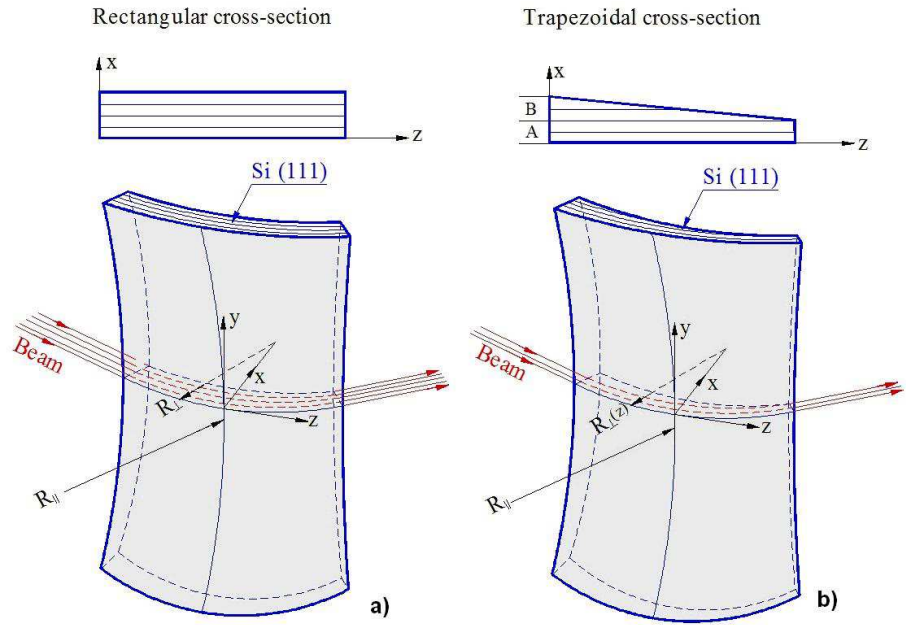}}
{\caption{The scheme  of a uniform bend of a crystal - strip of rectangular cross-section. b - a nonuniform bend of a plate of a trapezoidal  cross-section.}
\label{fig-1}}
\end{figure}

\begin{figure}
\scalebox{0.5}{\includegraphics{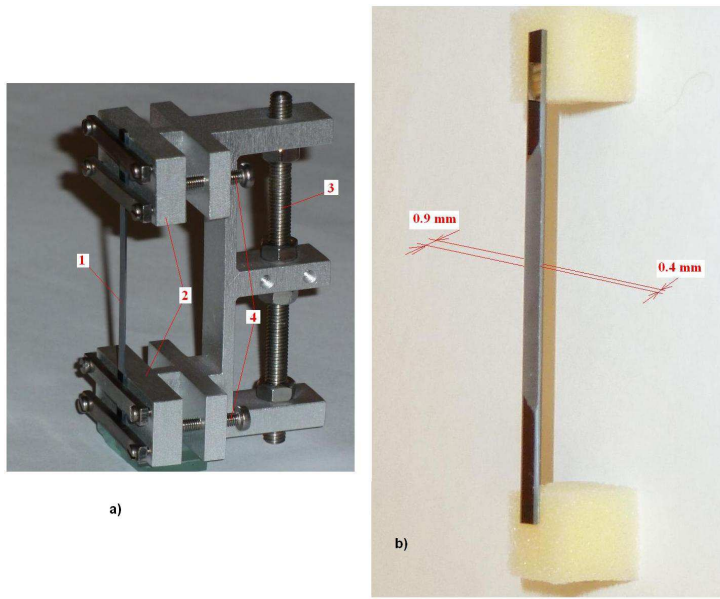}}
{\caption{a - photo of the crystal device, b -  view  of a crystal plate from the back face.
              }
\label{fig-2}}
\end{figure}

 The image of a crystal from the back side is shown in Fig. 2b,  where the processed oblique face of trapezoidal cross-section  is visible.
 The crystal plate which has been cut out initially lengthways crystalline  plane (111) in form of a parallelepiped with 
the sizes $ (x  y  z) = (0.9 \times 70 \times   3)$ mm, in the central part is processed with a slope on a trapeze with a size  $x_1 = 0.9$ mm 
at a forward end face up to $x_2 = 0.4$ mm at a back end face. 
The shape of a horizontal bend of a crystal (along coordinate z) was measured by the laser device using scheme  
described in (\cite{9}, page 86). Results of measurement of the bending shape  of a plate are displayed in Fig. 3. 

\begin{figure}
\scalebox{0.5}{\includegraphics{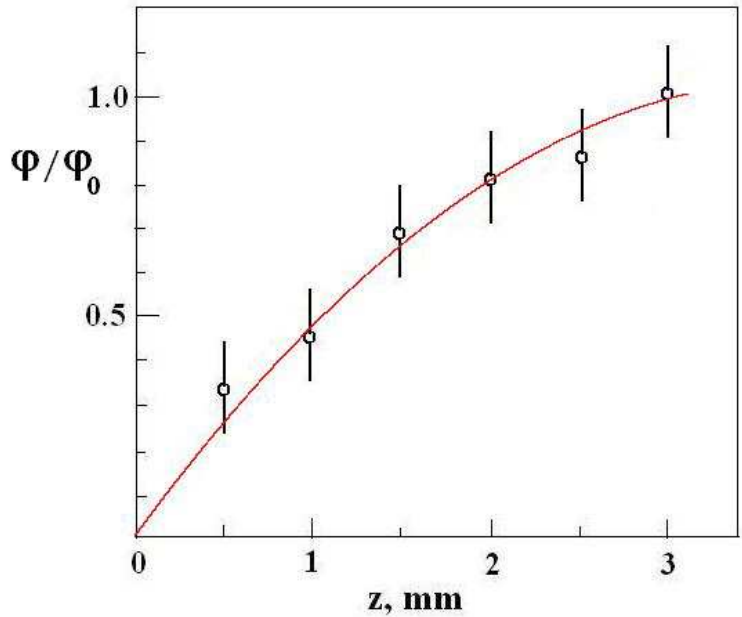}}
{\caption{Results of measurements of the bending angle as a function of $z$-coordinate (symbols) and its approximation with
the help of Eq. (6).
              }
\label{fig-3}}
\end{figure}

     The full angle  of a bend is equal $\varphi_0 = 0.3$ mrad. Changing of bending radius along coordinate z can be approximated
 by the dependence: 
\begin{equation}
R(z)={R_0 \over 1-Cz/L_0},
\end{equation} 
where $R_0$ is the bending radius at $z=0$, $L_0$ is the single crystal thickness  and $C= 0.8$ is the constant value.   
The calculated bending angle $\varphi (z)=\int_0^z dz/R(z)$ (see the curve in Fig. 3) is in a good agreement with
 the measurements.

The realized parameters of a crystal device approach for an optimum deflection  of particles with multi-GeV energy, experiments 
with which are carried out  in IHEP  and CERN \cite{11,12}. 
      It is necessary to note, because of  trapezoidal cross-section of a crystal, on a full angle  are bent only crystalline  planes marked on fig.1b  
as  area  A. The planes  shown as an area B  have the different bending angles  according to the length of the oblique side of a trapeze. 
At crystal extraction of a beam from accelerator  only the area A  of a crystal is involved in the process, as high energy 
particles will penetrate into a crystal (on coordinate $x$) on distance no more than 100 micron (for  the case of the LHC energy \cite{13}).

At the same time, the area B  of a crystal  can be used for focusing of the particle trajectories due to the  difference of  bending angles  
depending on coordinate x of incoming  particles. For the task of focusing the area  B can be  expand, changing the cross-section 
of a crystal up to  triangle as shown in fig.5. 
It is necessary to note, experiments on focusing of high energy particle trajectories by a crystal  with the oblique end face were 
carried out  intensively in the beginning of  90-ties \cite{14,15}. Novelty of application of the suggested crystal device consists in simplicity of a design 
and absence of superfluous substance around of a working crystal (devices \cite{14} had massive details around of a crystal). 
\section{ Examples of applications of the developed crystal device in particle accelerators }
\subsection{ Improvement of a crystal extraction/collimation of a circulating beam  }

   Fig. 4a shows the layout of the use of a short bent crystal for beam collimation in an accelerator. Particles of a circulating beam increase the amplitude of transverse oscillations due to numerous effects, such as scattering on a residual gas, the effect 
of nonlinearities, and the interactions at the point of contact. As a result, the beam halo where particles fall on the leading 
edge of the crystal appears. Due to the channeling effect, the majority of the beam halo (fraction 1) is 
 deeply deflected to the absorber. Only several percents of particles are deflected at an incomplete angle due to dechanneling 
(fraction 2), which leads to radiation losses on the accelerator (secondary particles 3). A crystal with decreasing curvature can reduce
 the fraction of dechanneled particles \cite{8}. The thing is that particles moving in the crystal are scattered from the electrons and nuclei 
of the lattice and, hence, some of them leave the channeling mode (dechanneling process). At the same time, due to an increase 
in the curvature radius along the direction of  particle penetration, the available channeling region expands \cite{8}, particles 
move away from the maximum scattering boundary, and the fraction of dechanneled particles decreases. 

\begin{figure}
\scalebox{0.55}{\includegraphics{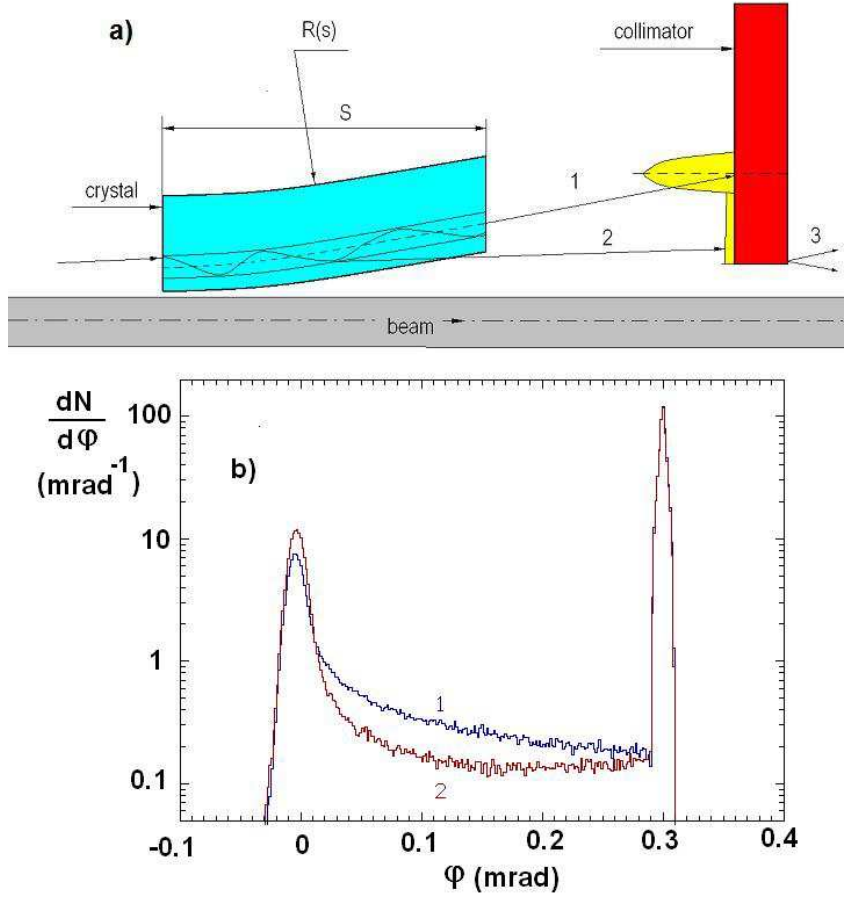}}
{\caption{A principle of beam collimation  with the help of a short channeling  crystal (a).  
 Distribution of 400 GeV protons  on deflection angles in 3 mm  Si (111) crystal  in the case of  constant 
curvature (1) and decreasing  curvature (2) (b). 
              }
\label{fig-4}}
\end{figure}

      That is, the fraction of the particles deflected  by a crystal on an incomplete angle 
 (it is less than  bend of a crystal, see Fig.4a) decreases, that reduces losses of particles in the accelerator and 
improves efficiency of collimation or extraction.  In Fig. 4b  calculations of 400 GeV protons deflection by crystals are resulted lead 
 by a method of Monte Carlo with use of the program described in \cite{8,16}. From the plots  it is visible, that the fraction  of "harmful" 
particles (between peaks of falling and fully-deflected particles)  rejected  in our crystal in few  times in comparison with a case 
of  uniform  bent crystal. Calculations show also, that efficiency of beam extraction from U-70  can be increased from 
85 \% up to 95 \% with new crystal device application.  By corresponding optimization of parameters of the crystal, the 
similar device can be applied to improvement of beam collimation in the LHC.

\subsection{Focusing of high energy  particle beams}
Presented in Refs. \cite{14,15} the consideration of  focusing problem is based on a geometrical description
of the process and valid only for single crystals with a constant curvature and with a special shape of skew cut of the exit face.
In this paper we  investigate the  problem of focusing of high energy beams for  a common case, in particular, 
for bent single crystals with a variable curvature.

\begin{figure}
\scalebox{0.45}{\includegraphics{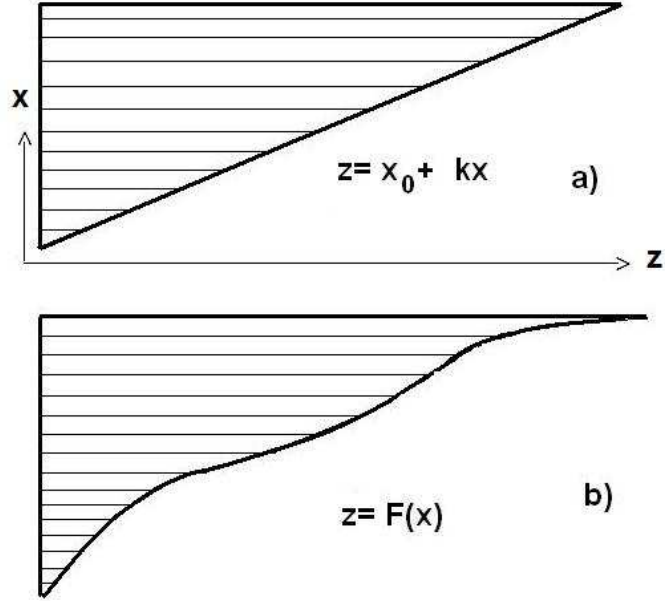}}
{\caption{Crystals with the linear (a) and  arbitrary (b) forms of cut.  Before usage they should be bent.
With the help of different bending methods  the bent crystals may be fabricated with a constant or variable curvature.
One can expect that the connection $z = F(x)$ is conserved in bent crystals. 
}
\label{fig-5}}
\end{figure}

The method for calculation of beam focusing is visual and simple. Let us consider the beam channeling in 
a bent single crystal of constant curvature  with skew cutting on the end (see, Fig. 5). On the exit from crystal
particles are deflected on the angle $\varphi= L/R + \theta$, where $R$ is the bending radius, $ L $ is
the path of particle in the body of a single crystal, and $\theta$ is the angle due to oscillation motion
of particles in a channeling regime. It is clear that an absolute value of $\theta$-angle is less than
the critical angle of channeling $\theta_c$.       
On the first stage we put $\theta=0$. Then in the case of single crystal without cut  all the particles
pass the length equal to $L$ and, hence, they are deflected on the same angle. At  the condition $\theta=0$ the beam 
after end of  a crystal is parallel. It is clear that for a single crystal with the cut  particles with
different transversal coordinates $x$ have different paths $L(x)$ and, hence, these particles have various
deflection angles. It turns out that at a special form of  the cut, particles focusing on some distance from
the crystal edge. We can write for the arbitrary form $F(x)$ of cut the following equation:
\begin{equation}
\sigma_x(l)= \int_0^d \rho(x) (X-\overline{X})^2 dx,
\end{equation}
where $\sigma_x(l)$ and $X = x+ \varphi l$ are the mean square size of beam and the particle coordinate
 on the distance $l$ from a single crystal, $\rho(x)$  is the normalized (on unit) distribution function over $x$-coordinate
(at $l=0$) and $\overline{X}=\int_0^d \rho(x) (x+ \varphi l)dx$.
From here, we get 
\begin{equation}
\sigma_x(l)= \langle x^2 \rangle -\overline{x}^2 +\langle \varphi^2 \rangle l^2 -\overline{\varphi}^2l^2
+2 \langle x \varphi \rangle l - 2\overline{x}\, \overline{\varphi} l.
\end{equation}
where $\langle x^2 \rangle$ and $\langle \varphi^2 \rangle $ are the mean square size of beam and the mean square deflection angle
 at $l=0$, $\overline{x}$ and $\overline{\varphi}$ are the mean size of beam and the mean deflection angle (at $l=0$) 
and $\langle x \varphi \rangle  = \int_0^d x \varphi(x) \rho(x) dx$ and 
$d$ is the transversal size of a single crystal. The $\sigma_x(l)$-function has a minimum when
\begin{equation}
l= l_f= -{\langle x\varphi \rangle - \overline{x} \, \overline{\varphi} \over \langle \varphi^2 \rangle - \overline{\varphi}^2}.
\end{equation} 
The focusing takes place at the condition $l_f >0$. It means that 
$\overline{x} \,\overline{\varphi} - \langle x \varphi \rangle > 0$. 
In the case when a crystal has a variable radius the deflection angle is defined by the relation:
\begin{equation}
\varphi(x)=\int_0^{F(x)}{dz\over R(z)}.
\end{equation}
Here  $z= F(x)$ is connection between $x$ and $z$ coordinates, which is determined by the form of cut.
Now we can take into account the natural divergence of a beam due to oscillator motion of particles in the channeling regime.
As result, we get for the total mean square size $\sigma_T(l)= \sigma_x(l)+ \langle \theta^2 \rangle l^2$.
We ignored the distribution over $x$ coordinate due to a small value of atomic interplanar distance.  This consideration is similar
to consideration of paper \cite{17}, where the sum distribution of particles may be represented as a convulsion of two
independent distributions. 

\begin{figure}
\scalebox{0.5}{\includegraphics{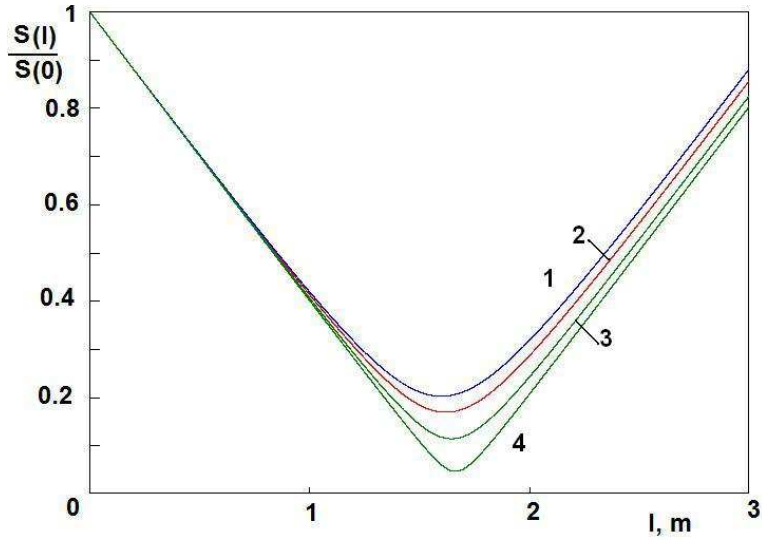}}
{\caption{The relative variation of square mean size of beam $(S(l)/S(0)=\sqrt{\sigma_x(l)/ \sigma_x(0)})$ as a function 
of distance $l$. For additional information, see  the text.
              }
\label{fig-6}}
\end{figure}

Let us apply our equations to a single crystal with increasing bending radius which presented in Fig. 1b.
The deflection angle for a single crystal with the variation of radius defined by Eq. (1) is  
\begin{equation}
\varphi(z)= {z\over R_0} - {Cz^2 \over 2L_0 R_0}.
\end{equation} 
In accordance with Fig.1b  we can take $z=F(x)=x_0+ kx$. Note that we can put $x_0=0$ by  a corresponding choice of the
coordinate system. In our further calculations we use the following quantities:
$C=0.8,L_0=0.3$ cm, $d=0.05$ cm, $k=-6$, $R_0=6$ m. Note that $\varphi (x)= ax+bx^2$, where $a=k/R_0$, $b=Ck^2/(2L_0R_0)$.
Taking into account  a small transversal size of  a crystal we take the flat distribution function over $x$-coordinate
$\rho(x)=1/d$. Then we get
$\overline{x}=d/2$, $\langle x^2 \rangle=d^2/3$, $\overline{\varphi}=ad/2 +bd^2/3$, 
$\langle \varphi^2 \rangle =a^2d^2/3 +abd^3/2 +b^2d^4/5$, $\langle x \varphi \rangle =ad^2/3 + bd^3/4$.   
Substitution of these quantities in Eq, (3) gives the envelope of a beam.
In this way we get  for $l_f$:
\begin{equation}
l_f= -{M_2\over M_1},
\end{equation}
and for the mean square beam size at $l=l_f$:
\begin{equation}
s_{min}={d^2\over 12} -{M_2^2\over M_1},
\end{equation}
where
\begin{equation}
M_1=[a^2d^2/12 +abd^3/6+4b^2d^4/45+\langle \theta ^2 \rangle],
\end{equation}
\begin{equation}
M_2=ad^2/12+bd^3/12.
\end{equation}
It is interesting that for the linear dependence $\varphi (x)= ax$,
\begin{equation}
l_f=-{1\over a+12\langle \theta^2 \rangle/(ad^2)},
\end{equation}
\begin{equation}
s_{min}={\langle \theta^2 \rangle/a^2 \over 1+ 12\langle \theta^2 \rangle/(a^2d^2)} .
\end{equation}
One can assume for estimations that  the distribution over $\theta$-angle is flat. It means that 
$\langle \theta ^2 \rangle \approx \theta_c^2/6$.

Fig. 6 illustrates the behavior of  square mean size of beam as a function of $l$ for above mention crystal parameters. 
The  curve 1 is calculated for $\theta_c =0$ and the  curve 2 is calculated for $\theta_c^2/6 = 10^{-10}$.
The last value corresponds approximately to a beam energy $\approx 70$ GeV. For energies $\sim 400$ GeV and more 
the both curves are very close in between. We see that $l_f \approx 1.62 $ m and the minimal size of beam is
in $\approx 5$ times less than initial one. Note that $l_f$-value is proportional to the bending radius.  

\begin{figure}
\scalebox{0.5}{\includegraphics{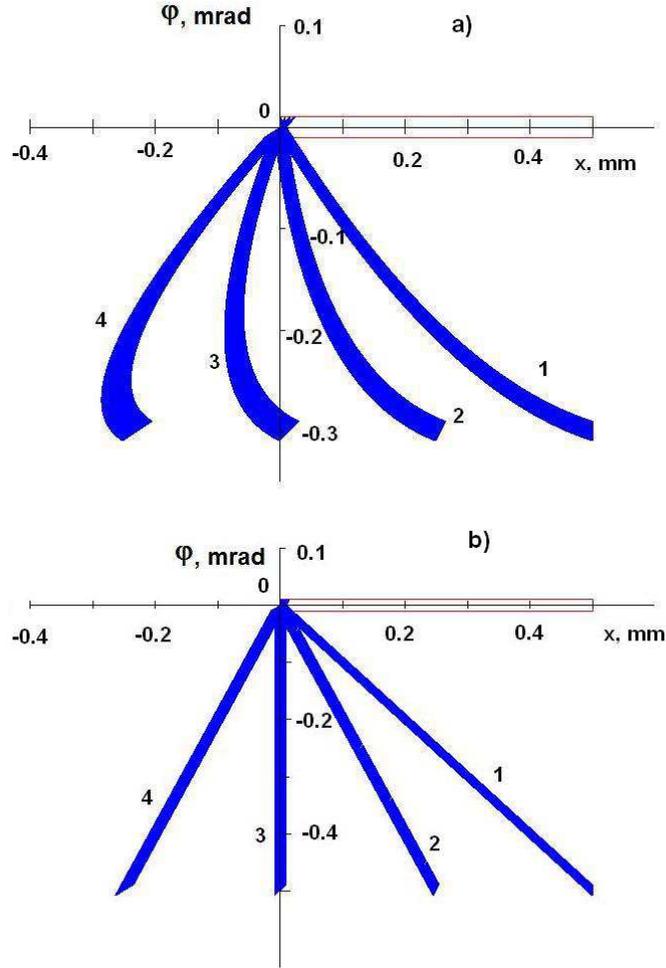}}
{\caption{ The phase volume of a beam at different l = 0 (the curve 1) , $l_f/2$ ( 2), $l_f$ (3). $1.5 l_f$ (4) .  The crystal parameters corresponds to Fig. 1b.
The crystal with the linear coordinate-angle couple (b).
The  rectangle is the accepted phase volume. $\theta_c=10 \mu$rad.
              }
\label{fig-7}}
\end{figure}

Fig. 7a  illustrates the transformation of the two dimensional beam phase volume at different $l$ for mention above crystal parameters. 
Fig. 7b illustrates the same volume but for the case when  $\varphi (x)= ax$ is a linear function, From comparison of these figures
it is easy to understand that for a linear angle-coordinate dependence the size of beam is less significantly, or, in other words,
in this case the focusing is optimal. 
The same fact
also follows from Eqs.(7)-(12).  However, the advantage of crystal under consideration (with increasing radius) in the comparison 
with the linear case is more small losses of particles due to dechanneling \cite{8}. 

In the papers \cite{14,15} a geometrical description of focusing (for crystals with a constant curvature) was considered.
Here for focusing the special shape of cut was created. This shape represents the crossing of two cylindrical surfaces with different radii ($R$ and $r$)
One radius (R) is equal to the bending radius of crystal.  However, it is easy to get that for  large enough radii and thicknesses of crystal
about 1 mm the couple between $x$ and $z$ coordinates is a  linear with high accuracy. 
Really, we can write
\begin{equation}
\tan \varphi = {R \over 2r\sqrt{1-R^2/(4r^2)}},
\end{equation}
For small thicknesses of a crystal and large enough radii we
can represent $R=R_m+ x$, where $R_m$ is the minimal radius of crystal bending.
Then we get the angle-coordinate couple:
\begin{equation}
\varphi (x) \approx {R_m +x \over 2r\sqrt{1-R^2/(4r^2)}}.   
\end{equation}

\begin{figure}
\scalebox{0.5}{\includegraphics{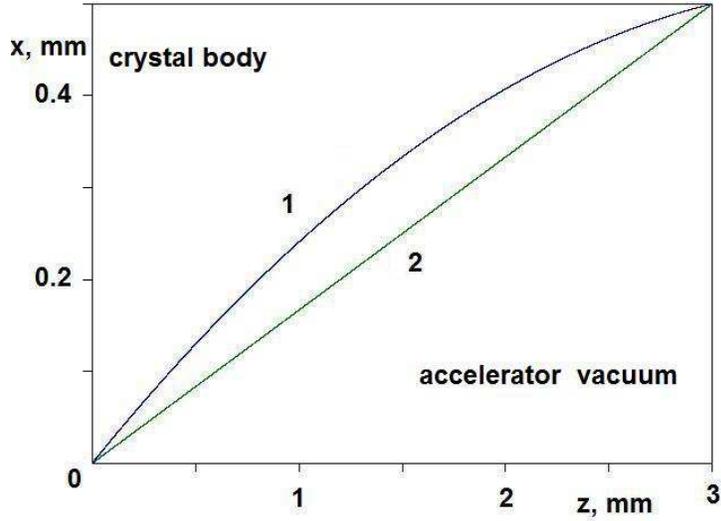}}
{\caption{Comparison of the improved  form of cut (the curve 1)  with linear one (the curve 2).
              }
\label{fig-8}}
\end{figure}

From our consideration follows that for a single crystal with varied curvature and  linear
function of the cut  $F(x)$ the focusing is not optimal. This situation may be corrected 
by creation of a special form of cut. Fig. 3 illustrates a good enough description of varied curvature
with the help of Eq. (5) which is consistent with measurements. It is easy with the help of Eqs. (4) and (5)
to find the corrected form 
\begin{equation}
z=F(x)={L_0 \over C}(1-\sqrt {1-{2CR_0\kappa  x \over L_0}}),
\end{equation}
where $\kappa =(1-C/2)L_0/(R_0d)$ is the coefficient in  $\varphi= \kappa x$. Obviously, the function defined
by Eq. (15) is  independent of the bending radius. Fig. 8 illustrates this function for parameters of a crystal under consideration in the paper.
The curves 3 and 4 (see Fig. 6) show  the envelope of beam  for corrected shape for 70 and 400 GeV protons, correspondingly. 

\begin{figure}
\scalebox{0.5}{\includegraphics{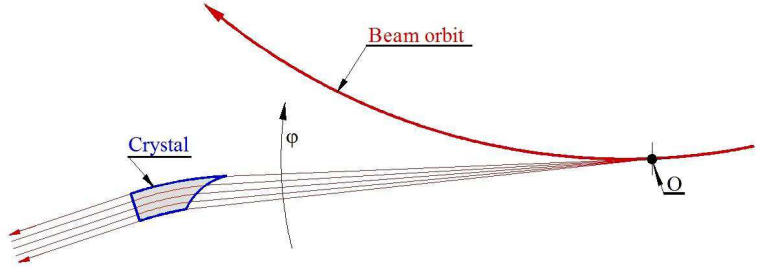}}
{\caption{The example of application of a focusing crystal for research of low-angular  processes.
              }
\label{fig-9}}
\end{figure}

     The focusing property  of the developed device can be applied on the LHC or the other accelerator of high energy 
to research of low-angular  processes. The crystal can be align on a fix  target by  focusing end face, as shown in Fig. 9. 
Rotating the crystal around of an axis O, one can  deflect the particles from the target  aside from adverse background 
area near the circulating beam. I.e. the role of a crystal consists in creation of clean conditions for registration of the necessary particles.
Other motive of such scheme  is production  of  the 
secondary particle beams in accelerators  by rather simple 
way.

\section{Conclusion}
      Developed and checked up optically  the crystal device is offered to be used in experiments with crystals on accelerators of IHEP  and CERN [11,12]. Two positive features characterize this device:

- the property of suppression of dechanneling for beam collimation improvement,

- the opportunity to focus the particle trajectories on small distances, which is useful tool for high energy physics.
\section{Acknowledgments}
      This work was supported by the Russian Foundation for Basic Research, project nos. $ 8\_02\_13533\_ofi\_ts$, $08\_02\_01453\_a$, and $11\_02\_90415\_Ukr\_f\_a$.

\end{document}